\documentclass{moriond}

\usepackage{amsmath}

\def\be{\begin{equation}}
\def\ee{\end{equation}}
\def\bea{\begin{eqnarray}}
\def\eea{\end{eqnarray}}

\newcommand{\GeV}{\ensuremath{\mathrm{GeV}}}
\newcommand{\MeV}{\ensuremath{\mathrm{MeV}}}
\newcommand{\GeVsq}{\ensuremath{\mathrm{GeV}^2}}
\newcommand{\Qsq}{\ensuremath{Q^2}}

\newcommand{\as}{\ensuremath{\alpha_s}}
\newcommand{\asmz}{\ensuremath{\alpha_s(M_Z)}}

\newcommand{\ptjet}{\ensuremath{P_{\rm T}^{\rm jet}}}

\newcommand{\etalab}{\ensuremath{\eta_{\rm lab}^{\rm jet}}}
\newcommand{\sredcc}{\ensuremath{\sigma_{\rm red}^{c\overline{c}}}}
\newcommand{\sredbb}{\ensuremath{\sigma_{\rm red}^{b\overline{b}}}}
\newcommand{\xbj}{\ensuremath{x_{\rm Bj}}}
\newcommand{\chisq}{\ensuremath{\chi^{2}}}

\newcommand{\Photo}{}

\begin{document}
\vspace*{3cm}
\title{QCD Results from HERA}
\author{D. BRITZGER for the H1 and ZEUS Collaborations }

\address{DESY, Notkestr. 85, 22607 Hamburg, Germany}

\maketitle\abstracts{
  New results on the measurements of the hadronic final state in
  neutral-current deep-inelastic scattering at HERA are presented.
  A combination of reduced charm and beauty cross sections is
  presented and the 
  masses of the heavy quarks are determined to
  $m_c=1290\,(^{+78}_{-53})\,\MeV$ and 
  $m_b=4049\,(^{+138}_{-118})\,\MeV$.
  The measurement of the  production of prompt photons accompanied by
  a jet provides a precise test of QCD predictions.
  Measurements of jet production cross sections are presented and
  compared for the first time to next-to-next-to-leading order
  predictions (NNLO).
  The strong coupling constant is determined from inclusive jet and
  dijet production cross sections using NNLO predictions
  to $\asmz=0.1157\,(6)_{\rm exp}\,(^{+31}_{-26})_{\rm th}$.
}

\section{Introduction}
At the HERA collider electrons or positrons were collided with
protons at a centre-of-mass energy of $319\,\GeV$. 
The two multi-purpose experiments H1 and ZEUS collected data until
2007 with an integrated luminosity of about $0.5\,{\rm fb}^{-1}$ per
experiment.
The exploration of the hadronic final state in neutral-current (NC)
DIS events, such as the study of jets, heavy quarks or photons produced
in these events, provides precise constraints on QCD parameters.

\section{Charm and Beauty Cross Sections}
\label{subsec:F2cc}
Multiple measurements of open charm and beauty production cross sections have been
performed by the H1 and ZEUS experiments during different data-taking
periods and using different tagging techniques. 
A previous combination~\cite{oldF2cc} of charm cross sections is 
extended, taking new data into account, and also beauty cross sections
are combined for the first time~\cite{charmbeauty}.
This provides a consistent data set of reduced charm and beauty cross
sections, \sredcc\ and \sredbb, in the kinematic range of photon
virtuality $2.5<\Qsq<2000\,\GeVsq$ and Bjoerken-$x$ of
$3\cdot10^{-5}<\xbj<5\cdot10^{-2}$.

\begin{figure}[tbhp]
\hfill
\begin{minipage}{0.48\linewidth}
\centerline{\includegraphics[width=0.99\linewidth]{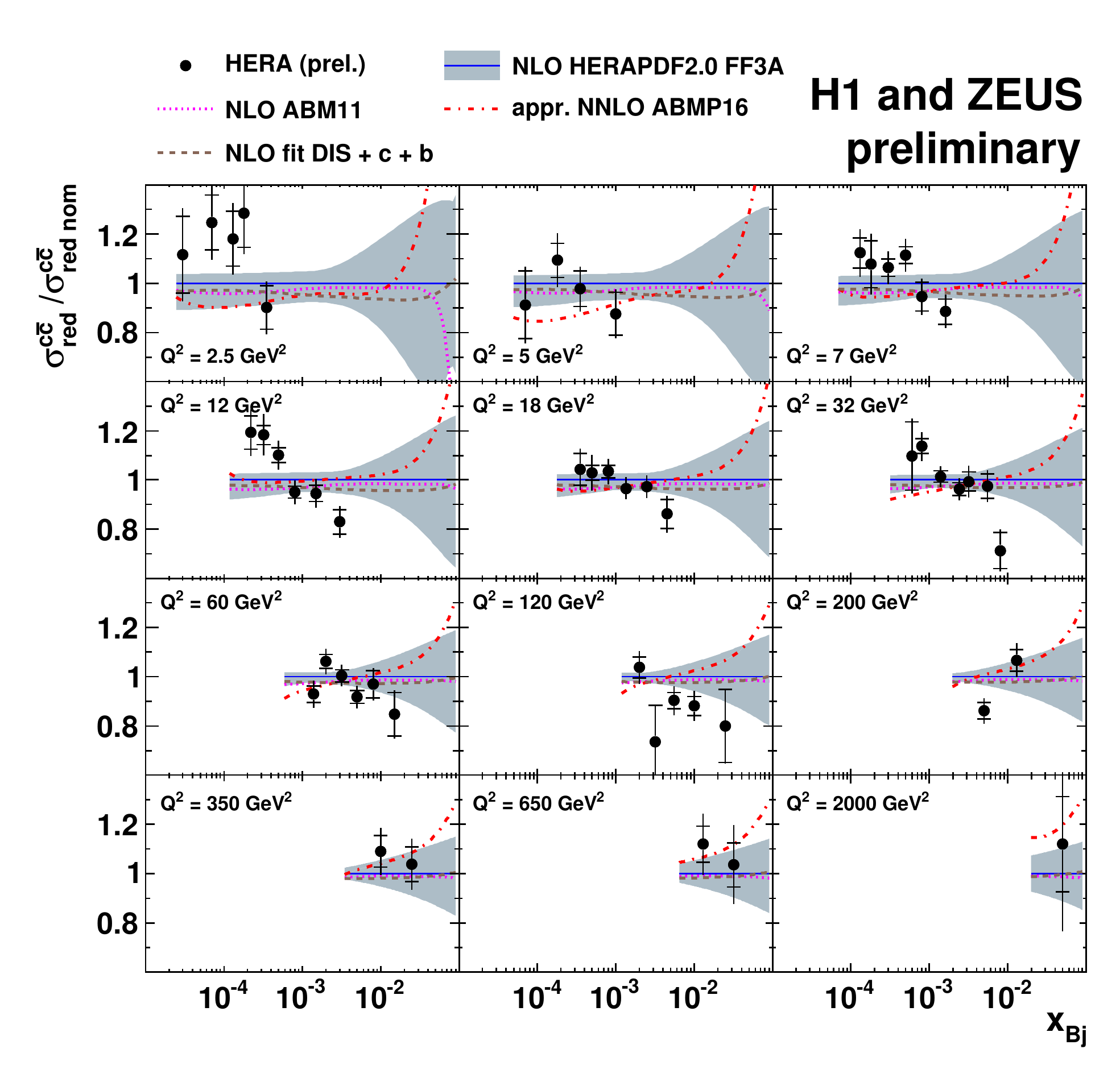}}
\end{minipage}
\hfill
\begin{minipage}{0.48\linewidth}
\centerline{\includegraphics[width=0.99\linewidth]{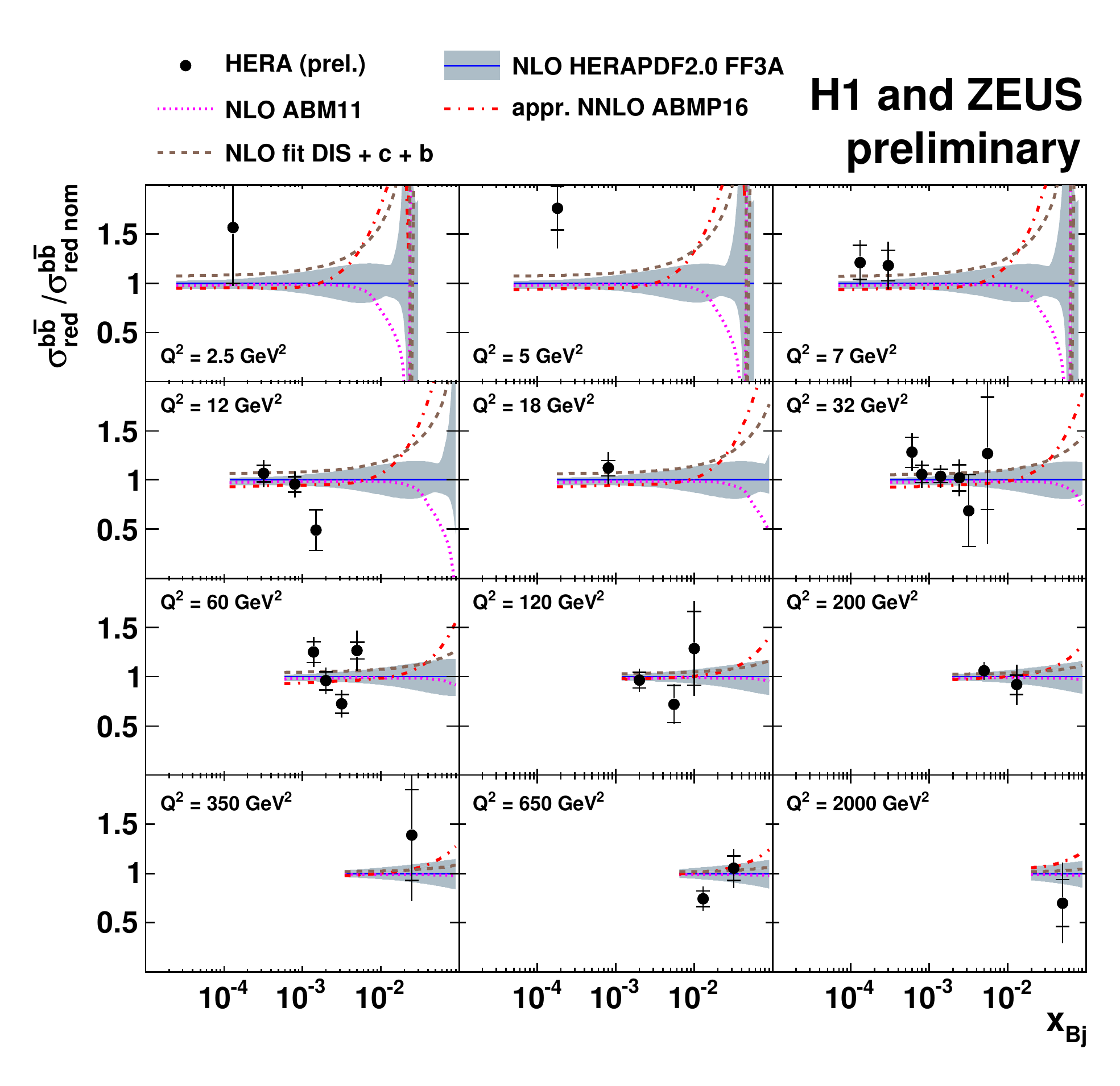}}
\end{minipage}
\hfill
\caption[]{Combined reduced charm (left) and beauty cross sections
  (right) as a function of \xbj\ for given values of \Qsq, displayed
  as ratios to NLO predictions using HERAPDF2.0FF3A.}
\label{fig:charmbeauty}
\end{figure}

The reduced cross sections which are input to the combination are
obtained by extrapolating the visible cross sections to the nearest
point of a (\xbj,\Qsq) grid using 
NLO predictions by the HVQDIS program.
The combination algorithm accounts for all correlations of the
uncertainties and in total 209 charm and 57 beauty measurements are
combined simultaneously to 52 reduced charm and 27 reduced beauty
cross sections.
The combination yields a value of $\chisq=149$ for 187 degrees of
freedom, 
thus indicating good consistency of the individual data. The
combined data have significantly reduced uncertainties.
The data are compared to NLO predictions using the HERAPDF2.0FF3A or
the ABM11 PDF sets, as well as to approximate NNLO predictions using
the ABMP16 PDF set in figure~\ref{fig:charmbeauty}. All predictions
are found to give a reasonable description of the data, with only
small differences between the studied predictions.

A QCD analysis of the combined charm and beauty reduced cross sections
together with the combined HERA inclusive DIS data~\cite{herapdf20} is
performed, where the predictions are
calculated by the QCDNUM and OPENQCDRAD programs in NLO accuracy.
The methodology follows closely the approach of
HERAPDF2.0FF3A~\cite{herapdf20}, employing the fixed-flavour scheme with
three active flavours at all scales, but in addition the masses of
the charm and beauty quarks are free parameters to the fit. 
In this fit the running charm and beauty quark masses are determined to
\begin{align}
 m_c(m_c) &= 1290\,({^{+46}_{-41}})_{\rm exp}\,({^{+62}_{-14}})_{\rm mod}\,({^{+7}_{-31}})_{\rm par}\,\MeV\,\,{\rm and} \\
\nonumber
 m_b(m_b) &= 4049\,({^{+104}_{-109}})_{\rm exp}\,({^{+90}_{-32}})_{\rm
   mod}\,({^{+1}_{-31}})_{\rm par}\,\MeV\,,
\nonumber
\end{align}
where the uncertainties indicate experimental uncertainties, model
uncertainties (mod), mainly dominated by scale variations by factors
of 2, and parameterisation uncertainties (par), which are determined using
fits with extended parameterisations of the PDFs. The inclusive DIS
data alone cannot reliably constrain the quark masses. The results are
consistent with other determinations.

\section{Isolated photons accompanied by jets in DIS}
\label{subsec:photons}
\begin{figure}
\hfill
\begin{minipage}{0.40\linewidth}
\centerline{\includegraphics[width=0.98\linewidth]{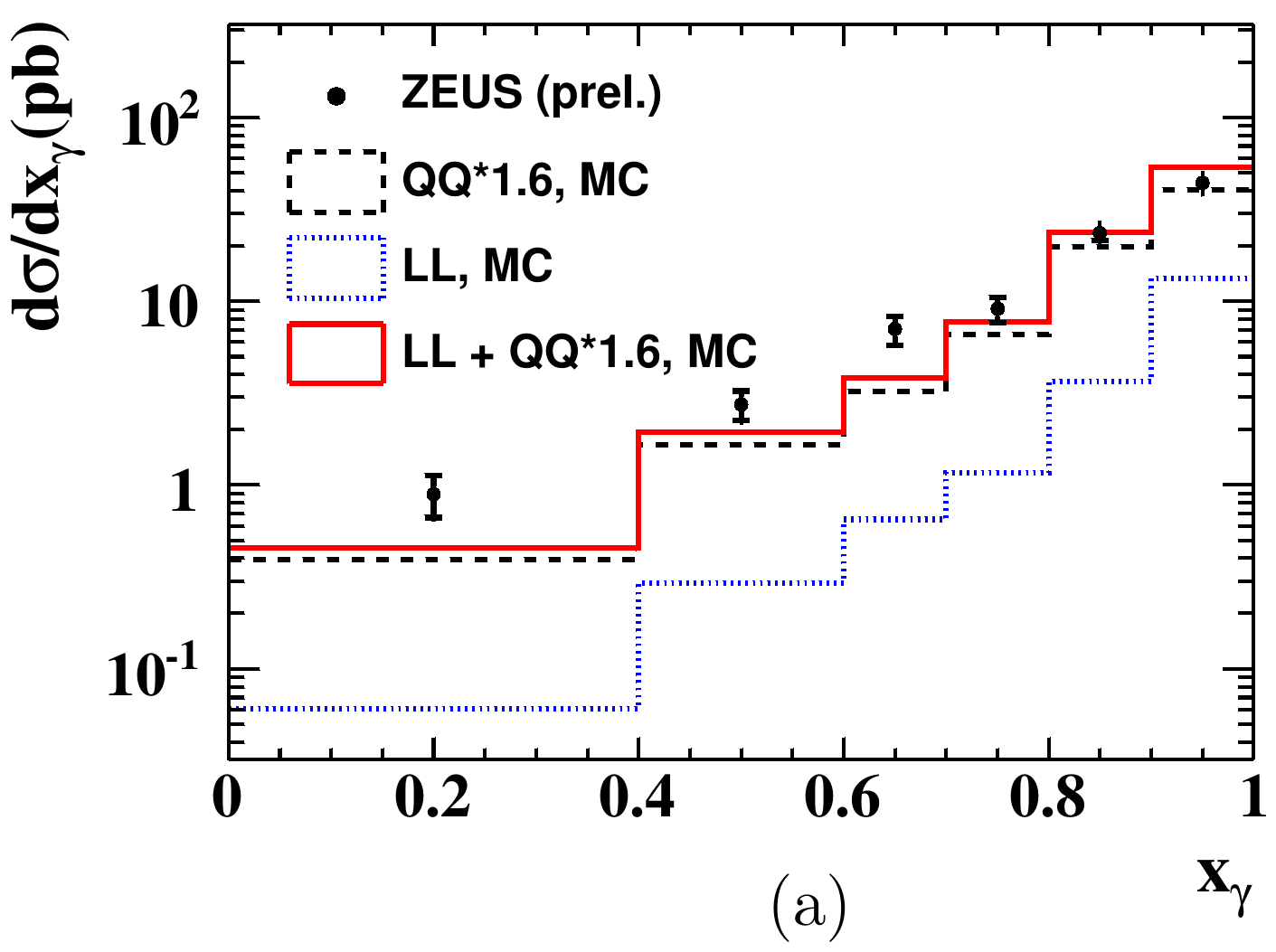}}
\caption[]{Differential cross sections for the production of photons
  accompanied by a jet in NC DIS as a function of fraction of the
  incoming photon energy that is given to the photon and
  the jet, $x_\gamma$, compared to MC
  predictions by Djangoh+Pythia.}
\label{fig:photons}
\end{minipage}
\hfill
\begin{minipage}{0.55\linewidth}
\centerline{\includegraphics[width=0.7\linewidth]{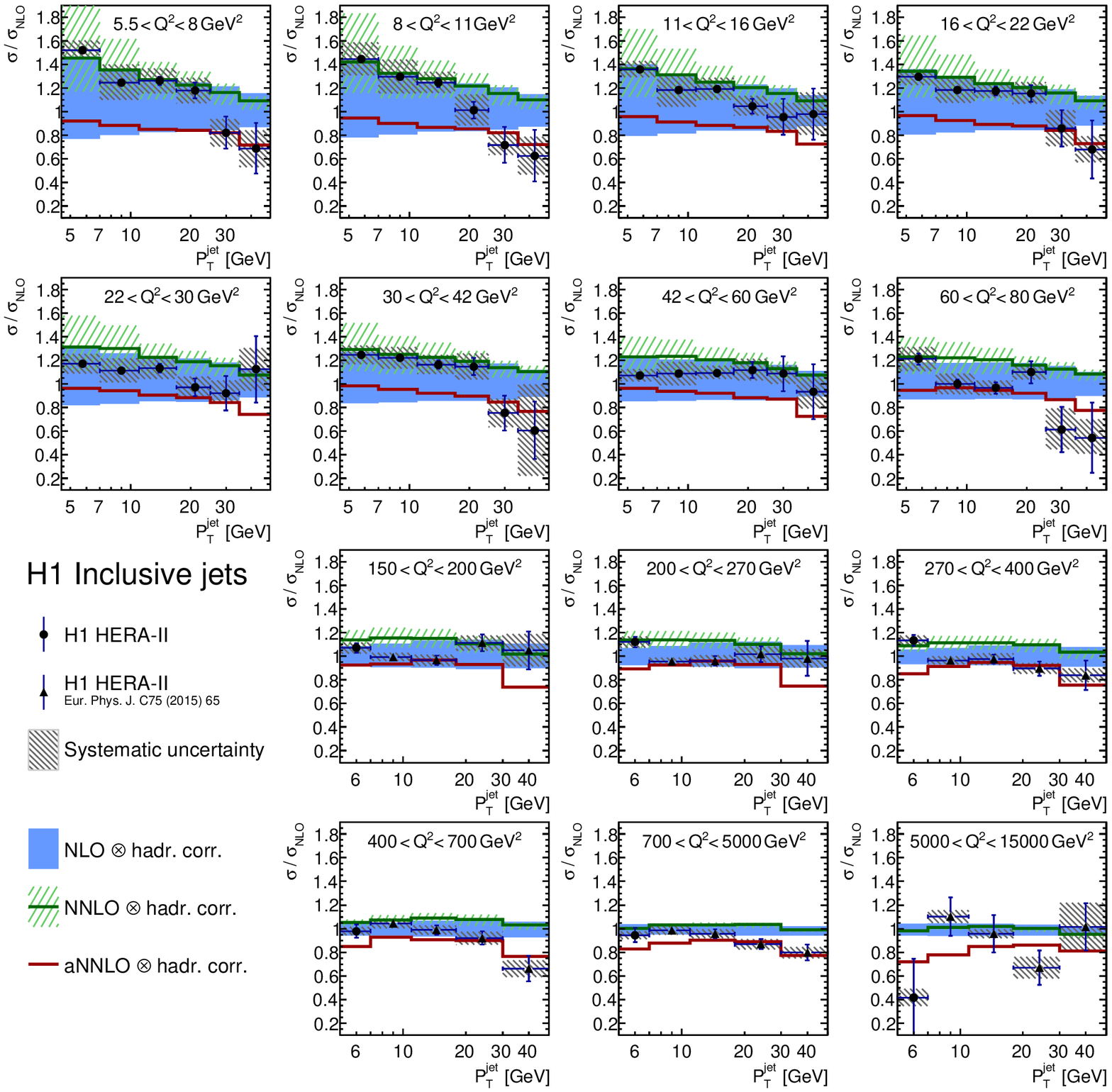}}
\caption[]{Inclusive jet cross sections in NC DIS in comparison to NLO,
  approximate NLO and full NNLO predictions.}
\label{fig:jets}
\end{minipage}
\hfill
\end{figure}

The production of isolated photons in events with at least one jet is
measured in NC DIS in the kinematic region of
$10<\Qsq<350\,\GeVsq$. Photons were measured with transverse energy
$4<E_T^\gamma<15\,\GeV$ and pseudorapidity $-0.7<\eta^\gamma<0.9$, and 
jets with $2.5<E_T^{\rm jet}<35\,\GeV$ and $-1.5<\eta^\gamma<1.8$. 
The analysis complements earlier measurements of
isolated photon production in NC DIS~\cite{photonsold}, and now differential
cross sections as functions $x_\gamma$, $x_p$, $\Delta\eta$,
$\Delta\phi$, $\Delta\eta_{e,\gamma}$ and $\Delta\phi_{e,\gamma}$ are
provided, where the observables denote the fraction of the
incoming photon energy that is given to the photon and
the jet, the fraction of proton energy taken by the interacting
parton, and the azimuthal angles
or rapidity differences between the photon and the jet or the
scattered lepton, respectively.
The cross sections are
compared to Monte Carlo predictions by Djangoh+Pythia, where the
photons radiated off a quark line are scaled by a factor of
1.6. These predictions provide a good description of the studied
distributions~\cite{photons}, as exemplarily displayed in
figure~\ref{fig:photons}. The cross sections are further compared to
predictions based on the $k_T$ factorisation method, which appear to
have some problems in describing all measured distributions equally well.

\section{Jet cross sections in DIS}
\label{subsec:jets}
Cross sections for jet production are measured in NC DIS in the Breit
 frame and exhibit a direct sensitivity to the strong coupling
constant and to the gluon content of the proton.
New measurements of jet production cross sections in NC DIS have been
performed in the kinematic region $5.5<\Qsq<80\,\GeV$ and
inelasticity $0.2<y<0.6$. These are inclusive jet cross sections
measured as a function of \Qsq\ and jet transverse momentum, \ptjet, as
well as dijet and trijet cross sections measured as functions of
\Qsq\ and the average \ptjet\ of the two or three leading jets~\cite{jets}.
Furthermore, the kinematic range of an earlier measurement of
inclusive jet cross sections~\cite{jetsold} at higher values of
\Qsq\ is extended to lower values of \ptjet.
The data are compared to NLO, to approximate NNLO and to 
next-to-next-to-leading order (NNLO) predictions~\cite{NNLO}, whenever available.
The ratio of inclusive jet cross sections to NLO
calculations together with other predictions is displayed in
figure~\ref{fig:jets}. 
The predictions are in general found in good agreement with the data
within the experimental and theoretical uncertainties.
The NNLO predictions improve significantly the description of
inclusive jet and dijet cross sections as compared to NLO predictions
in particular at lower scales and they also exhibit reduced
scale uncertainties.
Measurements of jet cross sections normalised to the
inclusive NC DIS cross section in the respective \Qsq\ interval
further improve the experimental precision, since experimental
uncertainties cancel partially.

\section{Strong coupling constant at NNLO from jet cross sections}
\label{subsec:alphas}
The strong coupling constant \asmz\ is determined in a fit of NNLO
predictions~\cite{NNLO} to inclusive jet and dijet cross sections measured by the
H1 experiment during different data taking periods and different
center-of-mass energies~\cite{alphas}. Altogether five data sets of
inclusive jet cross sections and four data sets of
dijet cross sections are considered and the data
covers a range of momentum transfer of
$5.5<\Qsq<15\,000\,\GeVsq$ and jet transverse momenta of
$\ptjet>4.5\,\GeV$.
Jets are defined by the $k_t$ jet-algorithm with a
parameter $R=1$ and are contained within the pseudorapidity range
$-1<\etalab<2.5$.
The \as-extraction methodology accounts for the \as-dependence of the
hard coefficients, as well as the \as-dependence of the PDFs~\cite{alphas}. The
latter is predicted by the factorisation theorem and it is taken as such into
account in the fit. This provides an increased sensitivity to
\as\ at lower scales, but a decreased sensitivity at higher scales.

\begin{figure}
\hfill
\begin{minipage}{0.42\linewidth}
\centerline{\includegraphics[width=0.8\linewidth]{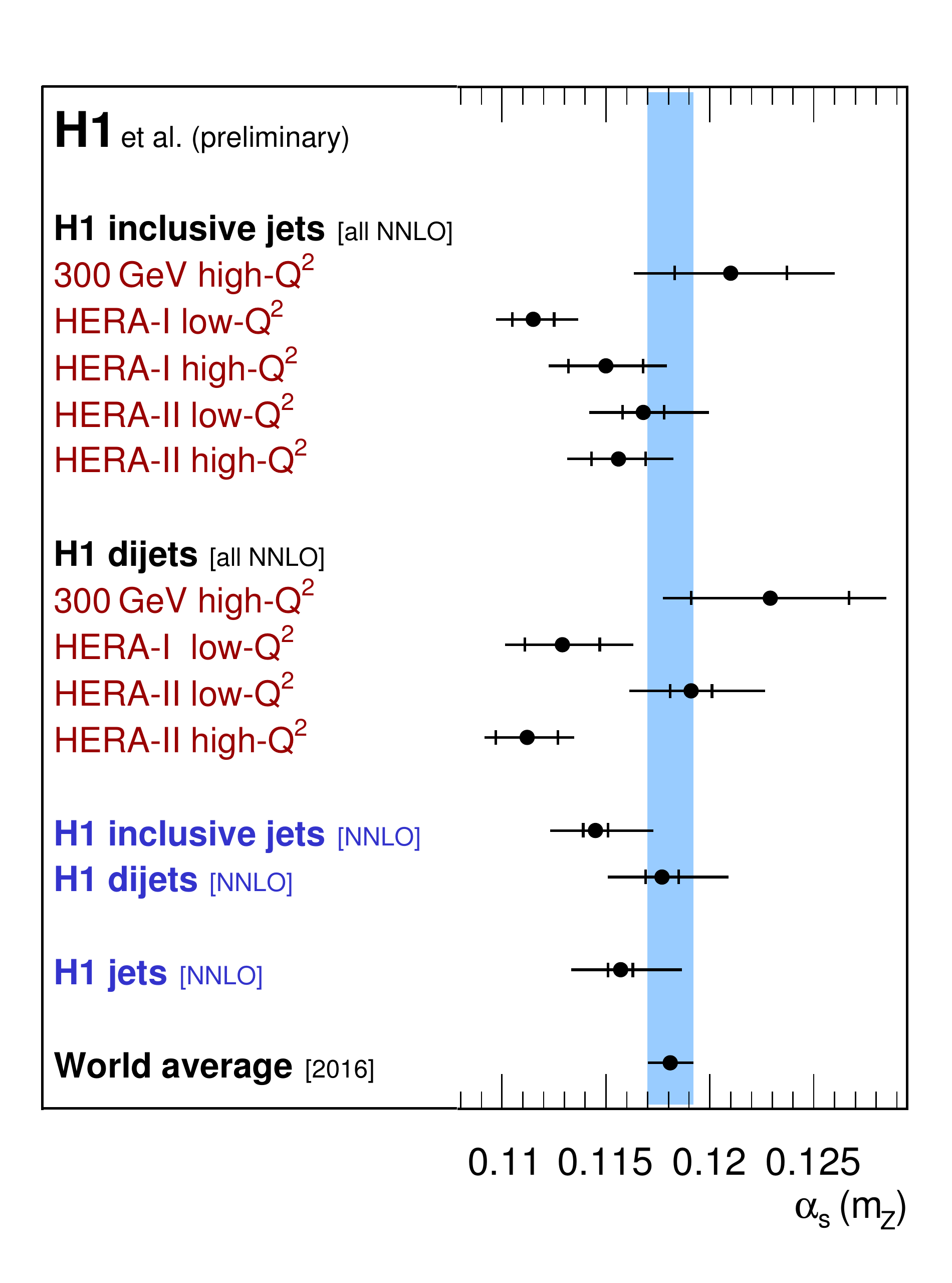}}
\caption[]{Values of the strong coupling constant, \asmz, determined in
  NNLO accuracy in fits to H1 jet cross section measurements. }
\label{fig:as1}
\end{minipage}
\hfill
\begin{minipage}{0.42\linewidth}
\centerline{\includegraphics[width=0.99\linewidth]{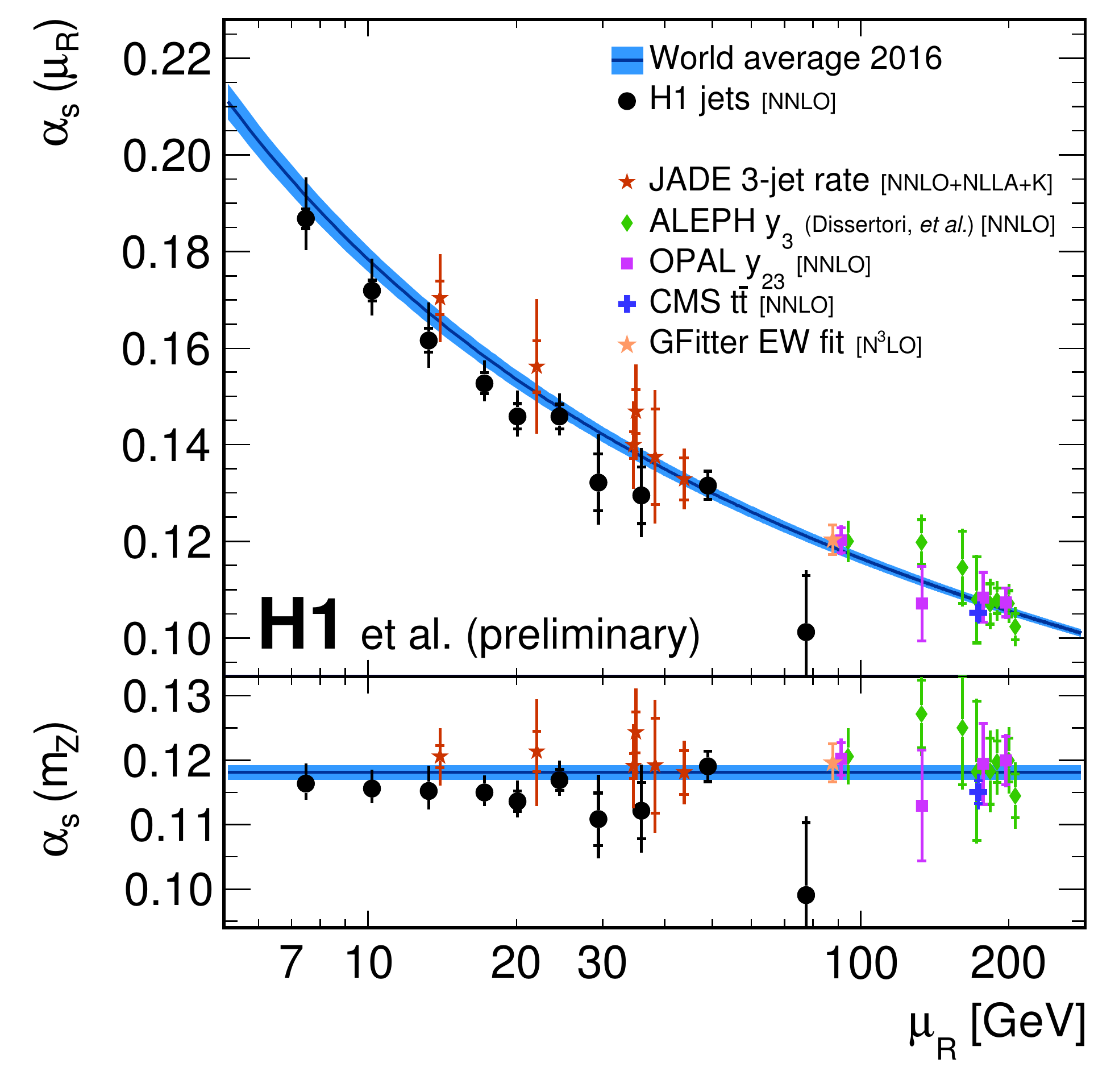}}
\caption[]{Results for \asmz\ in NNLO for fits to data points at similar
  scales in comparison to values from other experiments and processes.}
\label{fig:as2}
\end{minipage}
\hfill
\end{figure}

Values of \asmz\ are determined in fits to the individual data sets,
and in fits to all inclusive jet and dijet measurements as displayed in
figure~\ref{fig:as1}. These values are found to be consistent.
The value of \asmz\ determined in a fit to inclusive jet and dijet
cross sections is found to be
\begin{equation}
  \asmz = 0.1157\,(6)_{\rm exp}\, (3)_{\rm had}\, (6)_{\rm PDF}\,
  (12)_{\rm PDF\as}\, (2)_{\rm PDFset}\, (^{+27}_{-21})_{\rm scale}~,
  \nonumber
\end{equation}
where uncertainties on the PDF, the choice of the PDF set (PDFset),
the value of \asmz\ as input to the PDF extraction (PDF\as), the
uncertainty on the hadronisation correction, and uncertainties due to
scale variations by factors of 2, are considered.
The value of \asmz\ is consistent with the world average value of
$\asmz=0.1181\,(11)$. The dominating uncertainty arises from scale
variations of the NNLO predictions.

The running of the strong coupling constant is determined by repeating
the fits to groups of data points at similar scales. The results are
displayed in figure~\ref{fig:as2} and compared to other extractions in
at least NNLO accuracy. Good agreement with the QCD expectation and
other determinations is found.

\section{Summary and conclusions}
Several years after ending of data taking at the HERA accelerator
the H1 and ZEUS experiments have transformed their analysis frameworks
into a long-term usable computational environment and provide new
measurements and also combinations of previously published results.
New theoretical developments, such as full next-to-next-to-leading
order calculations for jet production, can thus be explored together with new
data and precision QCD results are obtained, such as precision
determinations of the strong coupling constant, \asmz, or the masses of
heavy quarks.





\section*{References}

\begin{thebibliography}{99}
 \bibitem{oldF2cc} H. Abramowicz et al. [H1 and ZEUS Collaborations],
   {\em Eur.Phys.J.} C73 (2013) 2, 2311, [arXiv:1211.1182].

 \bibitem{charmbeauty} H1 and ZEUS Collaborations, 
   H1 and ZEUS preliminary report, H1prelim-17-071, ZEUS-prel-17-01 (2017).

  \bibitem{herapdf20} H. Abramowicz et al. [H1 and ZEUS
    Collaborations], {\em Eur.Phys.J.} C75 (2015) 12, 580,
    [arXiv:1506.06042].
 
\bibitem{photonsold} 
  S. Chekanov et al. [ZEUS Collaboration], {\em Phys. Lett.} B687
  (2010) 16; \\
H. Abramowicz et al. [ZEUS Collaboration],
  {\em Phys.Lett.} B715 (2012) 88, [arXiv:1206.2270]. 

\bibitem{photons} ZEUS Collaboration, 
  ZEUS preliminary report, ZEUS-prel-16-001 (2016).

\bibitem{jets} V. Andreev {\emph et al.} [H1 Collaboration], \emph{Eur.Phys.J.} C77 (2017) 4, 215, [arxiv:1611.03421].

\bibitem{jetsold} V. Andreev {\emph et al.} [H1 Collaboration],
  \emph{Eur.Phys.J.} C75 (2015) 2, 65, [arxiv:1406.4709].

\bibitem{NNLO} 
  J. Currie  {\emph et al.}, {\em Phys.Rev.Lett.} 117 (2016) 4, 042001, [arXiv:1606.03991]; \\
  J. Currie  {\emph et al.}, [arXiv:1703.05977];  J. Niehues {\emph et al.}, these proceedings.

\bibitem{alphas} H1 Collaboration and V.~Bertone et al., 
  H1 preliminary report, H1prelim-17-031 (2017).


\end{thebibliography}
\end{document}